\documentclass{PoS}

\title{Dark Matter scenarios at IceCube}

\ShortTitle{Dark Matter scenarios at IceCube}

\author{\speaker{Marco Chianese}
         \thanks{I thank the organizers of the Neutrino Oscillation Workshop and the conveners of my session. I acknowledge the financial support by the Instituto Nazionale di Fisica Nucleare I.S. TASP and the PRIN 2012 ``Theoretical Astroparticle Physics" of the Italian Ministero dell'Istruzione, Universit\`a e Ricerca.}\\
        Dipartimento di Fisica {\it Ettore Pancini}, Universit\`a di Napoli {\it Federico II}, and INFN, 		        Sezione di Napoli, Complesso Universitario di Monte S. Angelo, I-80126 Napoli, Italy\\
        E-mail: \email{chianese@na.infn.it}}

\abstract{The recent study on the the 6-year up-going muon neutrinos by the IceCube Collaboration and the multi-messenger analyses support the hypothesis of a two-component scenario explaining the diffuse TeV--PeV neutrino flux. Depending on the steepness of the astrophysical power-law, an excess in the IceCube data is shown in the energy range 10--100~TeV (low-energy excess) or at PeV  (high-energy excess). In both cases, we characterize a two-component neutrino flux where decaying Dark Matter particles provide a contribution to the IceCube observations.}

\FullConference{Neutrino Oscillation Workshop\\
                 4 - 11 September, 2016\\
                 Otranto (Lecce, Italy)}

\begin{document}

\section{Introduction}

The recent discovery of a diffuse neutrino flux at the TeV--PeV range by the IceCube collaboration~\cite{IceCube} has ushered us into a new era for astroparticle physics. Due to the very feeble interaction with the other Standard Model particles, neutrinos represent the best messenger for observing and studying the cosmos. Indeed, the observation of extraterrestrial neutrinos provides an important tool that can be adopted to examine the acceleration mechanisms of hadronic cosmic-rays and the properties of both galactic and extragalactic astrophysical environments. Moreover, the IceCube Neutrino Telescope has measured the most energetic neutrinos, offering us the possibility to explore the neutrino physics at energies where phenomena beyond the Standard Model can be relevant.\\ \indent
The origin of the diffuse neutrino flux is still unknown. In the last few years, the scientific community has proposed a variety of astrophysical sources (extragalactic Supernovae and Hypernovae remnants~\cite{Chakraborty:2015sta}, blazars~\cite{Kalashev:2014vya}, and gamma-ray bursts~\cite{Waxman:1997ti}) as potential candidates for the IceCube observations. In general, under the reasonable assumption of a correlation with the spectrum of hadronic cosmic-rays, one expects that the differential neutrino flux has a power-law behavior
\begin{equation}
\frac{{\rm d}\phi^{\rm Astro}}{{\rm d}E_\nu {\rm d}\Omega} = \phi^{\rm Astro}_0 \left( \frac{E_\nu}{100~{\rm TeV}} \right)^{-\gamma}\,,
\label{eq:astro}
\end{equation}
where $\gamma$ is called {\it spectral index} and $\phi^{\rm Astro}_0$ is the normalization of the neutrino flux at 100~TeV. The astrophysical neutrino flux is assumed to be isotropic and it is characterized by an equal flavour ratio $\left(1:1:1\right)$ at the Earth. While the standard Fermi acceleration mechanism at shock fronts implies $\gamma = 2.0$ at first order, the spectral index can assume larger values depending on the astrophysical source considered. Indeed, in case of neutrinos produced by hadronuclear $p$--$p$ interactions then $\gamma \lesssim 2.2$~\cite{ppsources} (see also Ref.~\cite{Murase:2013rfa}), while for photohadronic $p$--$\gamma$ interactions the spectral index has to be larger than $2.3$~\cite{Winter:2013cla}.\\ \indent
The IceCube observations of different data samples have revealed a preference for a soft power-law spectrum and the IceCube combined analysis has provided  the best-fit $\gamma_{\rm best}=2.50\pm0.09$~\cite{IceCube}. However, very recently, the analysis on the 6-year up-going muon neutrinos shows that at high energy $\left(E_\nu \geq 100\,{\rm TeV}\right)$ the best-fit spectral index is $\gamma = 2.13$, value that is disfavoured by 3.3$\sigma$ with respect to $\gamma_{\rm best}$. As also stated by the IceCube collaboration, this tension may indicate the presence of a second galactic component in the diffuse neutrino flux. \\ \indent
The tension with the assumption of a single power-law is further strengthened by considering the {\it multi-messenger} analyses. Indeed, the contributions of different astrophysical sources to the IceCube spectrum are strongly constrained if one assumes a correlation between the diffuse neutrino flux and the isotropic diffuse gamma-ray background measured by Fermi-LAT~\cite{Ackermann:2014usa}. For instance, it has been pointed out that the contribution of star-forming galaxies ($p$--$p$ sources) has to be smaller than $\sim 30\%$ at $100$~TeV and $\sim 60\%$ at $1$~PeV~\cite{Bechtol:2015uqb} (see also Ref.~\cite{Murase:2015xka}). On the other hand, gamma-ray bursts~\cite{Aartsen:2014aqy} and blazars~\cite{blazarMulti} ($p$--$\gamma$ sources) can only provide a contribution of $\sim1\%$ and $\sim20\%$ to the IceCube neutrino spectrum, respectively. Therefore, the multi-messenger analyses suggest that a hard neutrino spectrum is preferred with respect to a soft one.
\begin{figure}[t!]
\centering
\includegraphics[width=0.42\textwidth]{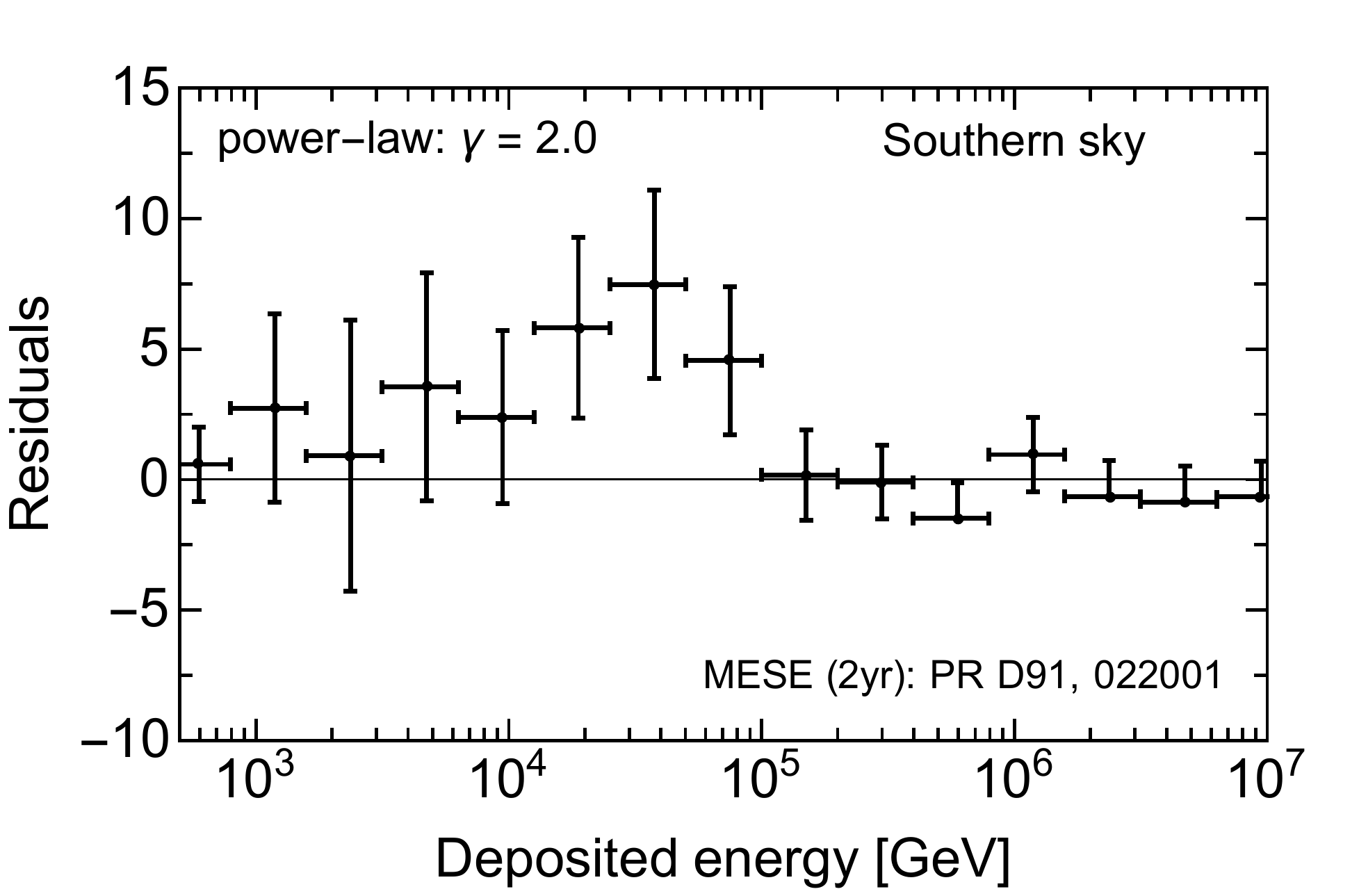}
\hskip3.mm
\includegraphics[width=0.42\textwidth]{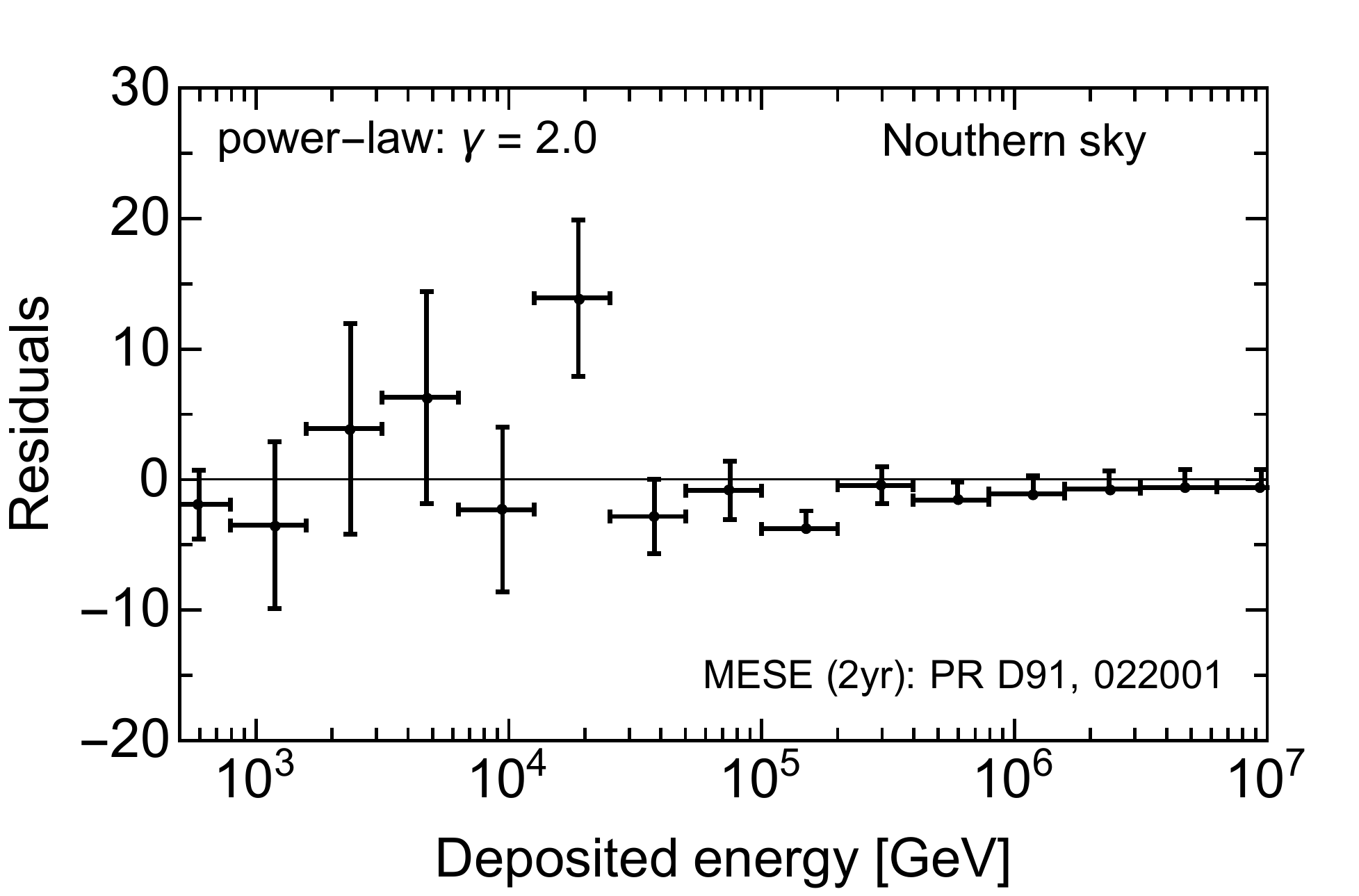}
\caption{\label{fig:1}Residuals in the number of neutrino events (2-year MESE) with respect to a single astrophysical power-law with spectral index 2.0, in the southern and northern hemispheres (see Ref.~\cite{Chianese:2016kpu}).}
\end{figure}\\ \indent
As discussed in Ref.s~\cite{Chianese:2016kpu,Chianese:2016opp}, once a hard neutrino spectrum $E_\nu^{-2.0}$ is considered according to the multi-messenger analyses, the IceCube data show an excess in the 10--100~TeV range ({\it low-energy excess}) with a maximum local statistical significance of 2.3$\sigma$ (2.0$\sigma$) in the 2-year MESE (4-year HESE) data (see Fig.~\ref{fig:1} for the residuals in the 2-year MESE data). However, as the spectral index increases, the excess moves towards PeV energies ({\it high-energy excess})~\cite{Boucenna:2015tra}.\\ \indent
All the above considerations lead to the conclusion that the IceCube measurements are explained in terms of a {\it two-component} neutrino flux. In particular, we have studied in detail the intriguing two-component scenario where, in addition to the neutrino background, one component is related to astrophysical sources and one is originated from Dark Matter~\cite{Chianese:2016kpu,Chianese:2016opp,Boucenna:2015tra} (see also Ref.~\cite{Chen:2014gxa}). In this case, the total differential neutrino flux takes the following expression
\begin{equation}
\frac{{\rm d}\phi}{{\rm d}E_\nu {\rm d}\Omega} = \frac{{\rm d}\phi^{\rm Astro}}{{\rm d}E_\nu {\rm d}\Omega} + \frac{{\rm d}\phi^{\rm DM}}{{\rm d}E_\nu {\rm d}\Omega}\,.
\label{eq:tot_flux}
\end{equation}
Here, the first term is given by the astrophysical power-law of Eq.~(\ref{eq:astro}). The Dark Matter second term (see Ref.~\cite{Chianese:2016kpu} for its detailed expression) depends on the particular interaction with the Standard Model particles (Dark Matter particles  decaying or annihilating into leptonic or hadronic final states) and on the halo density distribution of our galaxy.

\section{Two-component Neutrino Flux and Dark Matter}

\begin{figure}[t!]
\centering
\includegraphics[width=0.4\textwidth]{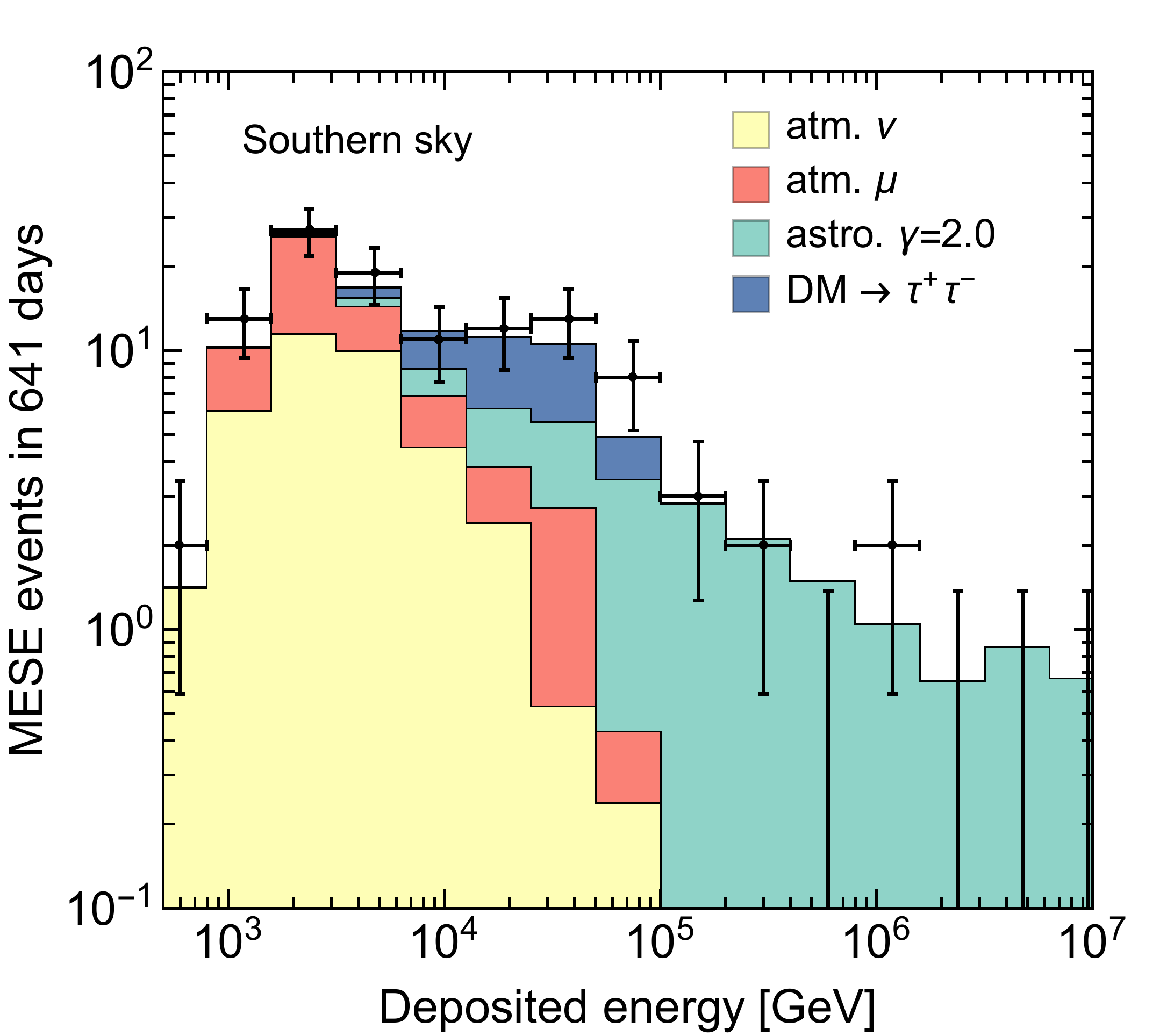}
\hskip3.mm
\includegraphics[width=0.4\textwidth]{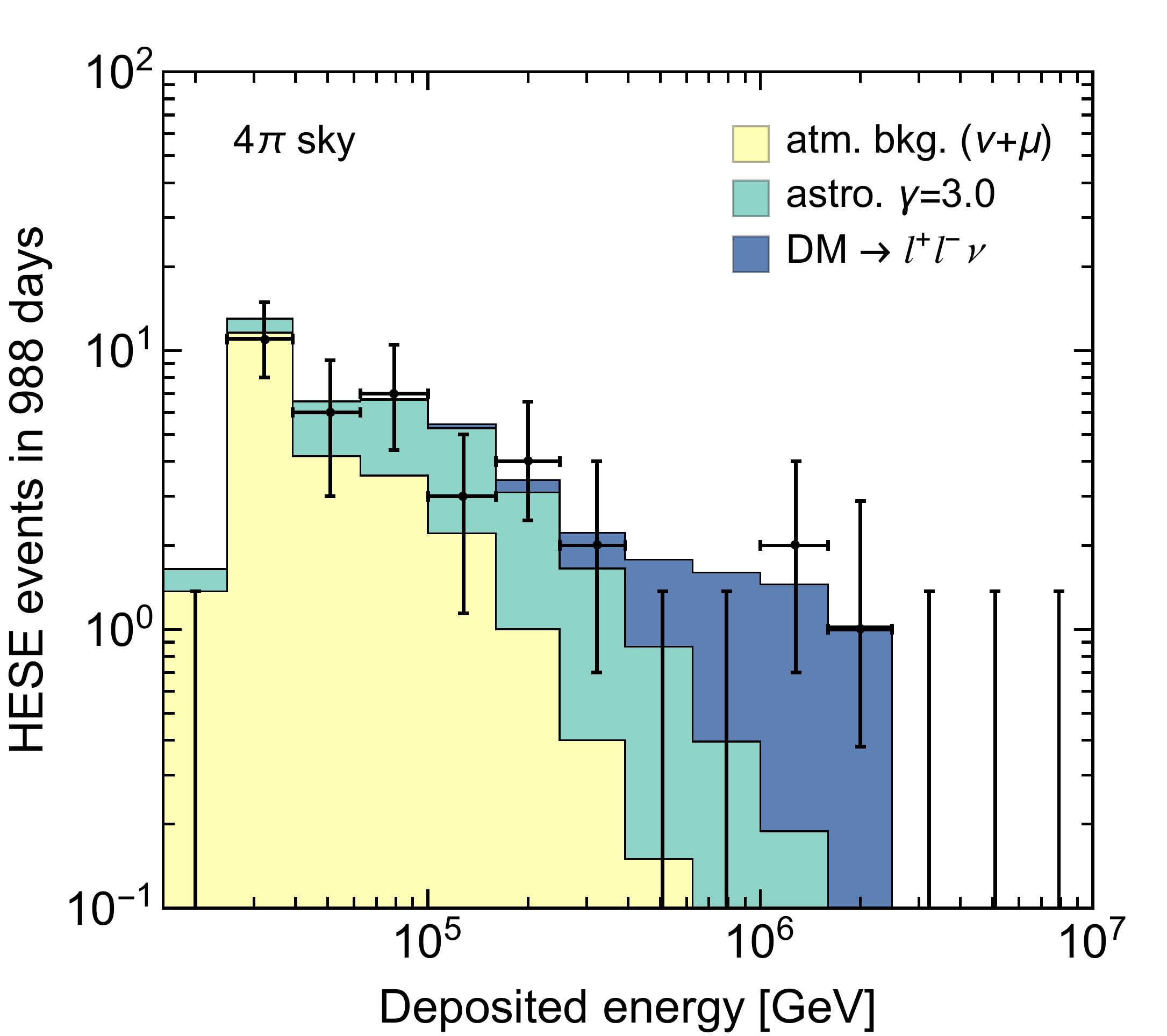}
\caption{\label{fig:2}Best-fit of the two-component hypothesis for the IceCube diffuse neutrino flux in case of 2-year MESE (left panel) and 3-year HESE (right panel) data (see Ref.s~\cite{Chianese:2016kpu}~and~\cite{Boucenna:2015tra}, respectively). The Dark Matter contribution is respectively obtained by considering Dark Matter decaying into tau leptons (low-energy excess) and leptophilic three-body decays at PeV energies (high-energy excess).}
\end{figure}

The proposed two-component interpretation of the IceCube data is depicted in left and right panels of Fig.~\ref{fig:2} for the low-energy excess~\cite{Chianese:2016kpu,Chianese:2016opp} and the high-energy one~\cite{Boucenna:2015tra}, respectively.\footnote{See references cited in Ref.s~\cite{Chianese:2016kpu,Chianese:2016opp,Boucenna:2015tra} for other Dark Matter interpretations of IceCube measurements.} The low-energy excess is explained in terms of the decay channel ${\rm DM} \to \tau^+\tau^-$, while the excess at high energy is fitted by the leptophilic model predicting ${\rm DM} \to \ell^+\ell^-\nu$. In both cases, the Navarro-Frenk-White halo density distribution has been considered. The low-energy excess has been characterized by analyzing the 2-year MESE data, since they have a lower energy threshold with respect to the HESE ones (see the different energy scale in the two plots). \\ \indent
Regarding the PeV neutrinos, the leptophilic model proposed in Ref.~\cite{Boucenna:2015tra} has two peculiar characteristics: the neutrino spectrum is spread differently from the case of two-body decays and it is peaked in a particular energy range due to the absence of quarks in the final states. Such a model is also able to account for the Dark Matter production in the early Universe through a freeze-in mechanism~\cite{Chianese:2016smc}. It is worth observing that considering the unrealistic behavior $E_\nu^{-3.0}$ of the astrophysical component is equivalent to assume a power-law with $\gamma=2.0$ exponentially suppressed for $E_\nu \geq 100$~TeV (neutrino spectrum expected for extragalactic Supernovae remnants~\cite{Chakraborty:2015sta}). 

\section{Conclusions}

The tension of the IceCube observations with the assumption of a single power-law flux indicates the presence of a second contribution to the diffuse neutrino flux. Therefore, we have examined the case where this second component is related to Dark Matter. A likelihood-ratio statistical test has shown that two-component scenario of Eq.~(\ref{eq:tot_flux}) is favoured by 2$\sigma$--4$\sigma$ with respect to a single power-law~\cite{Chianese:2016kpu}. The statistical significance depends on the Dark Matter model and the slope of the astrophysical contribution. In general, the leptophilic decaying Dark Matter models are preferred by both IceCube and Fermi-LAT measurements, while the other cases are disfavoured by multi-messenger studies. Better statistics could confirm the presence of an excess in the neutrino spectrum and this would potentially shed light on some of the deepest mysteries in contemporary physics: the nature of Dark Matter.


\begin{thebibliography}{99}
\bibitem{IceCube}
%\cite{Aartsen:2013jdh}
%\bibitem{Aartsen:2013jdh}
  M.~G.~Aartsen {\it et al.} [IceCube Collaboration],
  {\it Evidence for High-Energy Extraterrestrial Neutrinos at the IceCube Detector},
  Science {\bf 342} (2013) 1242856
  [\href{https://arxiv.org/abs/1311.5238}{\tt arXiv:1311.5238}].\\
  %\cite{Aartsen:2014gkd}
%\bibitem{Aartsen:2014gkd}
  M.~G.~Aartsen {\it et al.} [IceCube Collaboration],
  {\it Observation of High-Energy Astrophysical Neutrinos in Three Years of IceCube Data},
  Phys.\ Rev.\ Lett.\  {\bf 113} (2014) 101101
  [\href{https://arxiv.org/abs/1405.5303}{\tt arXiv:1405.5303}].\\
  %\cite{Aartsen:2014muf}
%\bibitem{Aartsen:2014muf}
  M.~G.~Aartsen {\it et al.} [IceCube Collaboration],
  {\it Atmospheric and astrophysical neutrinos above 1 TeV interacting in IceCube},
  Phys.\ Rev.\ D {\bf 91} (2015) no.2,  022001
  [\href{https://arxiv.org/abs/1410.1749}{\tt arXiv:1410.1749}].\\
  %\cite{Aartsen:2015knd}
%\bibitem{Aartsen:2015knd}
  M.~G.~Aartsen {\it et al.} [IceCube Collaboration],
  {\it A combined maximum-likelihood analysis of the high-energy astrophysical neutrino flux measured with IceCube},
  Astrophys.\ J.\  {\bf 809} (2015) no.1,  98
  [\href{https://arxiv.org/abs/1507.03991}{\tt arXiv:1507.03991}].\\
%\cite{Aartsen:2015zva}
%\bibitem{Aartsen:2015zva}
  M.~G.~Aartsen {\it et al.} [IceCube Collaboration],
  {\it The IceCube Neutrino Observatory - Contributions to ICRC 2015 Part II: Atmospheric and Astrophysical Diffuse Neutrino Searches of All Flavors},
  [\href{https://arxiv.org/abs/1510.05223}{\tt arXiv:1510.05223}].\\
%\cite{Aartsen:2016xlq}
%\bibitem{Aartsen:2016xlq}
  M.~G.~Aartsen {\it et al.} [IceCube Collaboration],
  {\it Observation and Characterization of a Cosmic Muon Neutrino Flux from the Northern Hemisphere using six years of IceCube data},
  [\href{https://arxiv.org/abs/1607.08006}{\tt arXiv:1607.08006}].
  
%\cite{Chakraborty:2015sta}
\bibitem{Chakraborty:2015sta}
  S.~Chakraborty and I.~Izaguirre,
  {\it Diffuse neutrinos from extragalactic supernova remnants: Dominating the 100 TeV IceCube flux},
  Phys.\ Lett.\ B {\bf 745} (2015) 35
  [\href{https://arxiv.org/abs/1501.02615}{\tt arXiv:1501.02615}].

%\cite{Kalashev:2014vya}
\bibitem{Kalashev:2014vya}
  O.~Kalashev, D.~Semikoz and I.~Tkachev,
  {\it Neutrinos in IceCube from active galactic nuclei},
  J.\ Exp.\ Theor.\ Phys.\  {\bf 120} (2015) no.3,  541
  [\href{https://arxiv.org/abs/1410.8124}{\tt arXiv:1410.8124}].

%\cite{Waxman:1997ti}
\bibitem{Waxman:1997ti}
  E.~Waxman and J.~N.~Bahcall,
  {\it High-energy neutrinos from cosmological gamma-ray burst fireballs},
  Phys.\ Rev.\ Lett.\  {\bf 78} (1997) 2292
  [\href{https://arxiv.org/abs/astro-ph/9701231}{\tt astro-ph/9701231}].
  
\bibitem{ppsources}
%\cite{Loeb:2006tw}
%\bibitem{Loeb:2006tw}
  A.~Loeb and E.~Waxman,
  {\it The Cumulative background of high energy neutrinos from starburst galaxies},
  JCAP {\bf 0605} (2006) 003
  [\href{https://arxiv.org/abs/astro-ph/0601695}{\tt astro-ph/0601695}].\\
  %\cite{Kelner:2006tc}
%\bibitem{Kelner:2006tc}
  S.~R.~Kelner, F.~A.~Aharonian and V.~V.~Bugayov,
  {\it Energy spectra of gamma-rays, electrons and neutrinos produced at proton-proton interactions in the very high energy regime},
  Phys.\ Rev.\ D {\bf 74} (2006) 034018
   Erratum: [Phys.\ Rev.\ D {\bf 79} (2009) 039901]
  [\href{https://arxiv.org/abs/astro-ph/0606058}{\tt astro-ph/0606058}].
  
%\cite{Murase:2013rfa}
\bibitem{Murase:2013rfa} 
  K.~Murase, M.~Ahlers and B.~C.~Lacki,
  {\it Testing the Hadronuclear Origin of PeV Neutrinos Observed with IceCube},
  Phys.\ Rev.\ D {\bf 88}, no. 12, 121301 (2013)
 [\href{http://arxiv.org/abs/1306.3417}{\tt arXiv:1306.3417}].

%\cite{Winter:2013cla}
\bibitem{Winter:2013cla} 
  W.~Winter,
  {\it Photohadronic Origin of the TeV-PeV Neutrinos Observed in IceCube},
  Phys.\ Rev.\ D {\bf 88}, 083007 (2013)
 [\href{http://arxiv.org/abs/1307.2793}{\tt arXiv:1307.2793}].
 
%\cite{Ackermann:2014usa}
\bibitem{Ackermann:2014usa}
  M.~Ackermann {\it et al.} [Fermi-LAT Collaboration],
  {\it The spectrum of isotropic diffuse gamma-ray emission between 100 MeV and 820 GeV},
  Astrophys.\ J.\  {\bf 799} (2015) 86
  [\href{https://arxiv.org/abs/1410.3696}{\tt arXiv:1410.3696}].
 
%\cite{Bechtol:2015uqb}
\bibitem{Bechtol:2015uqb}
  K.~Bechtol, M.~Ahlers, M.~Di Mauro, M.~Ajello and J.~Vandenbroucke,
  {\it Evidence against star-forming galaxies as the dominant source of IceCube neutrinos},
  [\href{https://arxiv.org/abs/1511.00688}{\tt arXiv:1511.00688}].
  
%\cite{Murase:2015xka}
\bibitem{Murase:2015xka}
  K.~Murase, D.~Guetta and M.~Ahlers,
  {\it Hidden Cosmic-Ray Accelerators as an Origin of TeV-PeV Cosmic Neutrinos},
  Phys.\ Rev.\ Lett.\  {\bf 116} (2016) no.7,  071101
  [\href{https://arxiv.org/abs/1509.00805}{\tt arXiv:1509.00805}].
  
%\cite{Aartsen:2014aqy}
\bibitem{Aartsen:2014aqy}
  M.~G.~Aartsen {\it et al.} [IceCube Collaboration],
  {\it Search for Prompt Neutrino Emission from Gamma-Ray Bursts with IceCube},
  Astrophys.\ J.\  {\bf 805} (2015) no.1,  L5
  [\href{https://arxiv.org/abs/1412.6510}{\tt arXiv:1412.6510}].

\bibitem{blazarMulti}
%\cite{Glusenkamp:2015jca}
%\bibitem{Glusenkamp:2015jca}
  T.~Gl\"usenkamp [IceCube Collaboration],
  {\it Analysis of the cumulative neutrino flux from Fermi-LAT blazar populations using 3 years of IceCube data},
  EPJ Web Conf.\  {\bf 121} (2016) 05006
  [\href{https://arxiv.org/abs/1502.03104}{\tt arXiv:1502.03104}].\\
  %\cite{Schimp:2015xha}
%\bibitem{Schimp:2015xha}
  M.~Schimp {\it et al.} [IceCube Collaboration],
  {\it Astrophysical interpretation of small-scale neutrino angular correlation searches with IceCube},
  PoS ICRC {\bf 2015} (2016) 1085
  [\href{https://arxiv.org/abs/1509.02980}{\tt arXiv:1509.02980}].\\
  %\cite{Aartsen:2016lir}
%\bibitem{Aartsen:2016lir}
  M.~G.~Aartsen {\it et al.} [IceCube Collaboration],
  {\it The contribution of Fermi-2LAC blazars to the diffuse TeV-PeV neutrino flux},
  [\href{https://arxiv.org/abs/1611.03874}{\tt arXiv:1611.03874}].

%\cite{Chianese:2016kpu}
\bibitem{Chianese:2016kpu}
  M.~Chianese, G.~Miele and S.~Morisi,
  {\it Dark Matter interpretation of low energy IceCube MESE excess},
    JCAP {\bf 1701} (2017) no.01,  007
  [\href{https://arxiv.org/abs/1610.04612}{\tt arXiv:1610.04612}].
  
%\cite{Chianese:2016opp}
\bibitem{Chianese:2016opp}
  M.~Chianese, G.~Miele, S.~Morisi and E.~Vitagliano,
  {\it Low energy IceCube data and a possible Dark Matter related excess},
  Phys.\ Lett.\ B {\bf 757} (2016) 251
  [\href{https://arxiv.org/abs/1601.02934}{\tt arXiv:1601.02934}].
  
%\cite{Boucenna:2015tra}
\bibitem{Boucenna:2015tra}
  S.~M.~Boucenna, M.~Chianese, G.~Mangano, G.~Miele, S.~Morisi, O.~Pisanti and E.~Vitagliano,
  {\it Decaying Leptophilic Dark Matter at IceCube},
  JCAP {\bf 1512} (2015) no.12,  055
  [\href{https://arxiv.org/abs/1507.01000}{\tt arXiv:1507.01000}].
  
%\cite{Chen:2014gxa}
\bibitem{Chen:2014gxa}
  C.~Y.~Chen, P.~S.~Bhupal Dev and A.~Soni,
  {\it Two-component flux explanation for the high energy neutrino events at IceCube},
  Phys.\ Rev.\ D {\bf 92} (2015) no.7,  073001
  [\href{https://arxiv.org/abs/1411.5658}{\tt arXiv:1411.5658}].
  
%\cite{Chianese:2016smc}
\bibitem{Chianese:2016smc}
  M.~Chianese and A.~Merle,
  {\it A Consistent Theory of Decaying Dark Matter Connecting IceCube to the Sesame Street},
  [\href{https://arxiv.org/abs/1607.05283}{\tt arXiv:1607.05283}].
 
\end{thebibliography}
\end{document}